\documentclass[conference]{IEEEtran}
\usepackage[T5,T1]{fontenc}
\usepackage[utf8]{inputenc}
\usepackage[vietnamese,english]{babel}
\usepackage{amsmath,amssymb,amsfonts}
\usepackage{algorithmic}
\usepackage{algorithm}
\usepackage{graphicx}
\usepackage{textcomp}
\usepackage{xcolor}
\usepackage{booktabs}
\usepackage{cite}
\usepackage{microtype}
\usepackage{tabularx}
\usepackage{url}
\usepackage{tikz}
\usetikzlibrary{shapes.geometric, arrows.meta, positioning, calc, shadows}

\begin{document}

\title{A Privacy-Preserving Edge AIoT Architecture for Multi-Dialect Vietnamese Speech Recognition and Early School Violence Detection}

\author{
\IEEEauthorblockN{
    Van Nguyen-Tran Lam\IEEEauthorrefmark{1}, 
    Phu-Gia Tran-Ha \IEEEauthorrefmark{1}, 
    Vinh Dang Ngoc\IEEEauthorrefmark{1},
    Thai Bao Huynh\IEEEauthorrefmark{2}, and
    Nghia Hieu Nguyen\IEEEauthorrefmark{2}
}
\IEEEauthorblockA{\IEEEauthorrefmark{1}Nguyen Hue High School, Tuy Hoa, Phu Yen, Vietnam\\
Email: \{lamvan070509, gia85965\}@gmail.com, dnvinh.thpt.nh@daklak.edu.vn}
\IEEEauthorblockA{\IEEEauthorrefmark{2}University of Information Technology, VNU-HCM, Ho Chi Minh City, Vietnam\\
Email: 23520105@gm.uit.edu.vn, nghiangh@uit.edu.vn}
}

\maketitle

\begin{abstract}
Preventing school violence in unsupervised and privacy-sensitive campus zones (such as corridors, stairwells, locker areas, and semi-private facilities) poses a fundamental Internet of Things (IoT) design dilemma: optical surveillance cameras are either legally prohibited or blind to verbal abuse, while cloud-centric audio IoT architectures violate student privacy, saturate local network bandwidth, and suffer from high round-trip latency. This paper presents an end-to-end, privacy-preserving \textbf{Edge AIoT} architecture for real-time Vietnamese Automatic Speech Recognition (ASR) and early school violence speech detection deployed directly on embedded edge hardware (NVIDIA Jetson Nano 4\,GB). To guarantee strict on-device privacy and computational efficiency, raw audio never leaves the edge node; instead, a cascaded acoustic gating pipeline combining Short-Time Energy (STE), Short-Time Fourier Transform (STFT) spectral filtering, and lightweight Voice Activity Detection (VAD) suppresses ambient school noise and triggers downstream neural inference only during sustained human speech. The linguistic engine integrates an FP16 TensorRT-optimized \textit{PhoWhisper-tiny} ASR model--enhanced via acoustic perturbation and sub-syllabic factorization to withstand phonological shifts and tonal mergers across Northern, Central, and Southern Vietnamese dialects--coupled with a contextual \textit{PhoBERT} classifier fine-tuned to distinguish benign peer banter from verbal threats, bullying, extortion, and distress calls. Only compact, anonymized MQTT alert payloads are transmitted over an isolated campus IoT network. Experimental evaluations demonstrate $524.5$\,ms end-to-end edge latency, a $10.7\%$ absolute Word Error Rate (WER) reduction on Central Vietnamese dialect speech, over $99.9\%$ IoT network bandwidth reduction compared to raw audio streaming, and a $92.4\%$ macro F1-score in school violence speech detection.
\end{abstract}


\section{Introduction}
School bullying, verbal coercion, extortion, and physical altercations remain critical safety threats in secondary education environments. Field evidence indicates that physical violence on school campuses rarely occurs instantaneously; rather, it is almost always preceded by audible precursors including hostile insults, cornering intimidation, extortion demands, or explicit victim pleas for help. However, deploying automated safety monitoring across campus blind spots---such as stairwells, hallways, locker rooms, and semi-private facilities where video cameras are ethically unacceptable or legally barred---presents severe architectural challenges for conventional Internet of Things (IoT) systems.

Traditional cloud-based audio IoT deployments stream continuous microphone signals from distributed edge sensors to centralized servers for transcription and analysis. In a school environment, this paradigm fails on three fronts: (1) transmitting continuous raw conversations violates student and staff privacy mandates; (2) streaming uncompressed $16$\,kHz PCM audio across dozens of campus IoT nodes consumes substantial network bandwidth ($256$\,kbps per node); and (3) cloud round-trip latency and network jitter delay timely intervention during rapidly escalating altercations. Consequently, shifting intelligence directly onto embedded edge devices (\textbf{Edge AIoT}) is essential.

Designing an effective Edge AIoT system for Vietnamese school violence speech detection introduces three core technical hurdles:
\begin{enumerate}
    \item \textbf{Edge Resource Constraints and Indoor Acoustic Reverberation:} Campus corridors and enclosed rooms exhibit strong acoustic reverberation alongside non-stationary background disturbances (footsteps, slamming doors, ventilation airflow, and distant crowd chatter). Continuously executing dense Transformer-based Automatic Speech Recognition (ASR) models on resource-constrained embedded hardware such as the NVIDIA Jetson Nano 4\,GB~\cite{nvidia2021jetson} rapidly causes thermal throttling, out-of-memory (OOM) failures, and unnecessary power consumption.
    \item \textbf{Multi-Dialectal and Phonological Complexity of Spoken Vietnamese:} Vietnamese is a tonal, analytical, monosyllabic language characterized by pronounced phonological divergence across Northern, Central, and Southern dialect regions. Under high emotional stress during confrontations, students speak in dense regional vernacular featuring initial consonant shifts (e.g., standard $/v/$ realized as glide $/j/$, $/r/$ shifting to $/z/$ or $/j/$), coda stop and nasal neutralizations (e.g., dental $/t/$ merging into velar $/k/$, alveolar $/n/$ shifting to velar $/\eta/$), regional slang, and tonal contractions (specifically in Central and Southern dialects where the broken tone \textit{\foreignlanguage{vietnamese}{ngã}} merges into the asking tone \textit{\foreignlanguage{vietnamese}{hỏi}}). Off-the-shelf ASR models trained predominantly on formal Northern speech suffer severe transcription degradation under these conditions.
    \item \textbf{Contextual Ambiguity in School Violence Language:} Adolescent verbal interactions are highly contextual. Explicit profanity frequently appears in friendly peer banter without malicious intent, whereas genuine extortion and bullying are often phrased in calm or superficially polite registers (e.g., coercing a student to hand over lunch money under threat of after-school retaliation). Simple keyword spotting or profanity blocklists produce high false-positive and false-negative rates.
\end{enumerate}

To address these challenges, this paper proposes a dedicated \textbf{Edge AIoT} framework tailored for Vietnamese campus violence prevention. Our main contributions are summarized as follows:
\begin{itemize}
    \item \textbf{Privacy-by-Design Edge AIoT Topology:} We design a decoupled edge-to-broker IoT architecture where raw audio is strictly confined to a volatile circular buffer in edge RAM, processed locally via cascaded spectral-VAD gating, and converted into lightweight, anonymized MQTT alert events with zero audio persistence.
    \item \textbf{Dialect-Resilient Vietnamese ASR on Embedded Hardware:} We adapt \textit{PhoWhisper-tiny}~\cite{le2024phowhisper} using acoustic perturbation, SpecAugment~\cite{park2019specaugment}, and sub-syllabic factorization (Onset--Nucleus--Coda)~\cite{nguyen2026atomictokensfactorizingsyllables}, compiling the model into an FP16 TensorRT engine capable of sub-$420$\,ms inference on the Jetson Nano.
    \item \textbf{Contextual School Violence Speech Classification:} We fine-tune \textit{PhoBERT}~\cite{nguyen2020phobert} with Focal Loss~\cite{lin2017focal} on domain-specific campus altercation dialogues integrated with Vietnamese toxic speech corpora~\cite{luu2021vihsd,hoang2023vihos,nguyen2021victsd} to accurately detect threats, extortion, and distress calls.
\end{itemize}

\section{Edge AIoT System Architecture and Operational Topology}
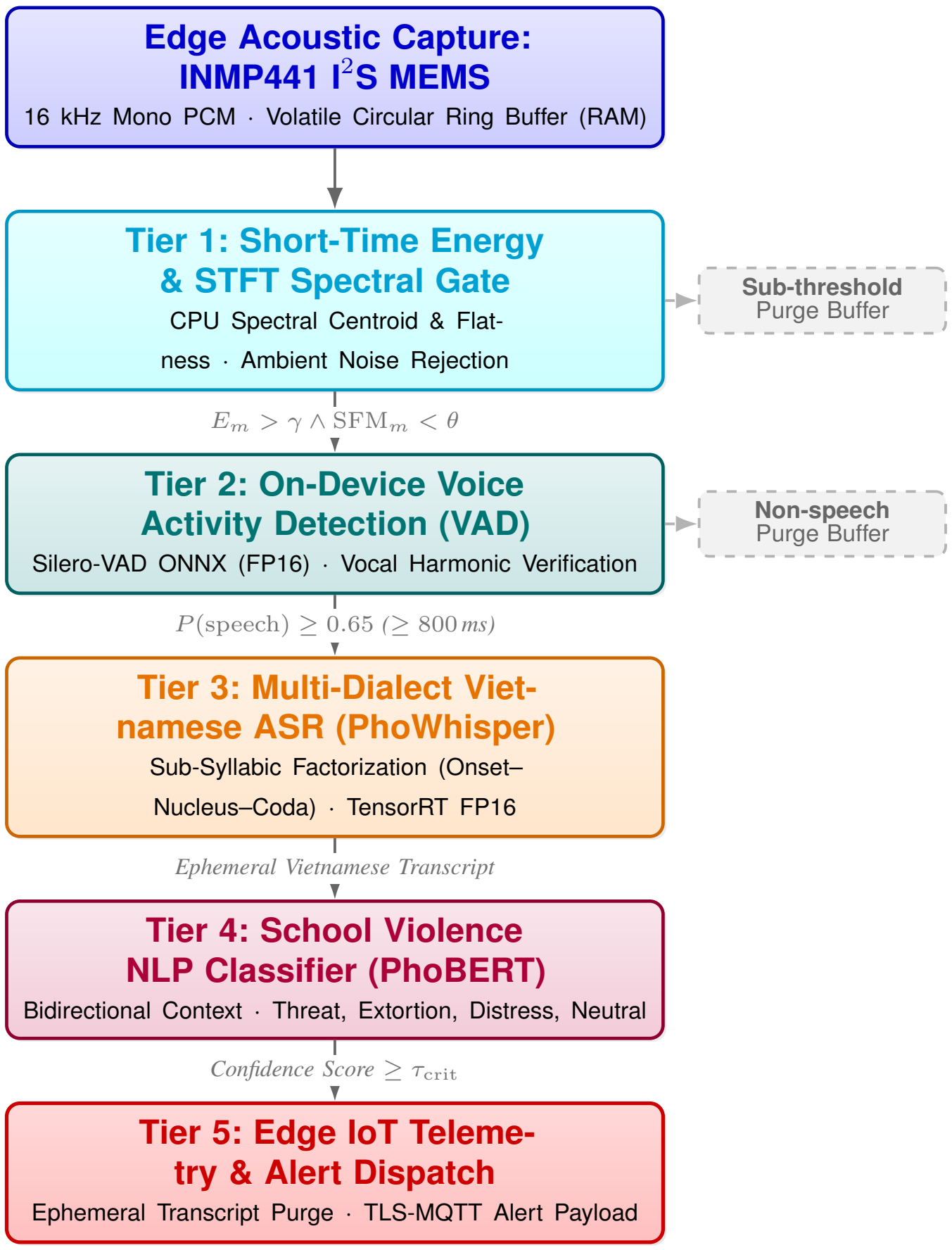
\begin{figure}[htbp]
\centering
\resizebox{\columnwidth}{!}{%
\begin{tikzpicture}[
    >=Latex,
    node distance=0.65cm and 0.4cm,
    font=\sffamily,
    % Box styles
    input_node/.style={
        rectangle, rounded corners=4pt,
        draw=blue!70!black, line width=1pt,
        top color=blue!10, bottom color=blue!20,
        text width=6.6cm, align=center,
        inner sep=5pt, drop shadow={opacity=0.15, shadow xshift=1pt, shadow yshift=-1pt}
    },
    tier1_node/.style={
        rectangle, rounded corners=4pt,
        draw=cyan!75!black, line width=1pt,
        top color=cyan!10, bottom color=cyan!20,
        text width=6.6cm, align=center,
        inner sep=5pt, drop shadow={opacity=0.15, shadow xshift=1pt, shadow yshift=-1pt}
    },
    tier2_node/.style={
        rectangle, rounded corners=4pt,
        draw=teal!75!black, line width=1pt,
        top color=teal!10, bottom color=teal!20,
        text width=6.6cm, align=center,
        inner sep=5pt, drop shadow={opacity=0.15, shadow xshift=1pt, shadow yshift=-1pt}
    },
    tier3_node/.style={
        rectangle, rounded corners=4pt,
        draw=orange!80!black, line width=1pt,
        top color=orange!10, bottom color=orange!20,
        text width=6.6cm, align=center,
        inner sep=5pt, drop shadow={opacity=0.15, shadow xshift=1pt, shadow yshift=-1pt}
    },
    tier4_node/.style={
        rectangle, rounded corners=4pt,
        draw=purple!75!black, line width=1pt,
        top color=purple!10, bottom color=purple!20,
        text width=6.6cm, align=center,
        inner sep=5pt, drop shadow={opacity=0.15, shadow xshift=1pt, shadow yshift=-1pt}
    },
    output_node/.style={
        rectangle, rounded corners=4pt,
        draw=red!80!black, line width=1.1pt,
        top color=red!15, bottom color=red!25,
        text width=6.6cm, align=center,
        inner sep=5pt, drop shadow={opacity=0.18, shadow xshift=1pt, shadow yshift=-1pt}
    },
    discard_node/.style={
        rectangle, rounded corners=3pt,
        draw=gray!60, line width=0.8pt, dashed,
        fill=gray!10, text=gray!80!black,
        text width=2.4cm, align=center, font=\scriptsize\sffamily,
        inner sep=3pt
    },
    edge_label/.style={
        font=\scriptsize\itshape, fill=white, inner sep=1.5pt, text=gray!90!black
    }
]

% Nodes Placement
\node[input_node] (in) {
    \textbf{\textcolor{blue!80!black}{Edge Acoustic Capture: INMP441 $\text{I}^2\text{S}$ MEMS}}\\[-1pt]
    {\scriptsize 16 kHz Mono PCM $\cdot$ Volatile Circular Ring Buffer (RAM)}
};

\node[tier1_node, below=of in] (t1) {
    \textbf{\textcolor{cyan!85!black}{Tier 1: Short-Time Energy \& STFT Spectral Gate}}\\[-1pt]
    {\scriptsize CPU Spectral Centroid \& Flatness $\cdot$ Ambient Noise Rejection}
};

\node[discard_node, right=0.35cm of t1] (d1) {
    \textbf{Sub-threshold}\\[-1pt]Purge Buffer
};

\node[tier2_node, below=of t1] (t2) {
    \textbf{\textcolor{teal!90!black}{Tier 2: On-Device Voice Activity Detection (VAD)}}\\[-1pt]
    {\scriptsize Silero-VAD ONNX (FP16) $\cdot$ Vocal Harmonic Verification}
};

\node[discard_node, right=0.35cm of t2] (d2) {
    \textbf{Non-speech}\\[-1pt]Purge Buffer
};

\node[tier3_node, below=of t2] (t3) {
    \textbf{\textcolor{orange!90!black}{Tier 3: Multi-Dialect Vietnamese ASR (PhoWhisper)}}\\[-1pt]
    {\scriptsize Sub-Syllabic Factorization (Onset--Nucleus--Coda) $\cdot$ TensorRT FP16}
};

\node[tier4_node, below=of t3] (t4) {
    \textbf{\textcolor{purple!90!black}{Tier 4: School Violence NLP Classifier (PhoBERT)}}\\[-1pt]
    {\scriptsize Bidirectional Context $\cdot$ Threat, Extortion, Distress, Neutral}
};

\node[output_node, below=of t4] (out) {
    \textbf{\textcolor{red!80!black}{Tier 5: Edge IoT Telemetry \& Alert Dispatch}}\\[-1pt]
    {\scriptsize Ephemeral Transcript Purge $\cdot$ TLS-MQTT Alert Payload}
};

% Connecting Arrows & Conditions
\draw[->, thick, gray!80!black] (in) -- (t1);

\draw[->, thick, gray!80!black] (t1) -- node[edge_label] {$E_m > \gamma \land \mathrm{SFM}_m < \theta$} (t2);
\draw[->, thick, gray!60, dashed] (t1) -- (d1);

\draw[->, thick, gray!80!black] (t2) -- node[edge_label] {$P(\mathrm{speech}) \ge 0.65$ ($\ge 800$\,ms)} (t3);
\draw[->, thick, gray!60, dashed] (t2) -- (d2);

\draw[->, thick, gray!80!black] (t3) -- node[edge_label] {Ephemeral Vietnamese Transcript} (t4);

\draw[->, thick, gray!80!black] (t4) -- node[edge_label] {Confidence Score $\ge \tau_{\mathrm{crit}}$} (out);

\end{tikzpicture}%
}
\caption{Cascaded privacy-preserving Edge AIoT architecture for multi-dialect Vietnamese speech recognition and school violence detection.}
\label{fig:architecture}
\end{figure}

The proposed system is architected as a self-contained Edge AIoT node coupled to an isolated campus telemetry plane, completely decoupling real-time audio signal processing from downstream IoT alerting. Fig.~\ref{fig:architecture} illustrates the end-to-end processing pipeline executed on the edge device.

The Edge AIoT architecture comprises three tightly integrated layers:
\begin{itemize}
    \item \textbf{Edge Acoustic Acquisition Front-End:} An omnidirectional digital MEMS microphone (INMP441) streams $16$\,kHz, 16-bit mono PCM audio directly over a hardware $\text{I}^2\text{S}$ bus into an NVIDIA Jetson Nano 4\,GB developer kit~\cite{nvidia2021jetson} (128-core Maxwell GPU, Quad-core ARM Cortex-A57 CPU). Incoming frames are written exclusively to a volatile circular Ring Buffer ($T = 3.0$\,s) in system RAM, ensuring zero non-volatile disk I/O.
    \item \textbf{Cascaded On-Device AI Inference Engine:} To maximize energy efficiency and prevent GPU thermal saturation on the Jetson Nano, heavy Transformer models remain idle during background silence or non-vocal noise. A lightweight CPU-based Short-Time Energy (STE) and STFT spectral gate first filters out ambient acoustic disturbances, followed by an FP16 Silero-VAD~\cite{silero2021vad} engine. Only verified human speech segments activate the TensorRT-accelerated \textit{PhoWhisper-tiny}~\cite{le2024phowhisper} ASR and \textit{PhoBERT}~\cite{nguyen2020phobert} violence intent classifier.
    \item \textbf{Lightweight IoT Telemetry and Alert Plane:} Rather than acting as a passive audio streamer, the edge node operates as an intelligent IoT publisher. Upon detecting verbal aggression above a critical confidence threshold $\tau_{\mathrm{crit}}$, the edge daemon immediately flushes the intermediate text transcript from RAM and publishes a compact JSON telemetry message ($< 256$\,bytes containing Node ID, Zone label, Incident Category, Timestamp, and Confidence Score) over an isolated, TLS-secured local MQTT broker to campus security responders.
\end{itemize}

\section{Edge-Native Privacy-Preserving Acoustic Gating}
In continuous IoT acoustic monitoring, running neural ASR models unconditionally wastes over $90\%$ of embedded compute cycles on background silence and non-speech events. Our edge daemon implements a deterministic two-stage acoustic gate to ensure that downstream GPU inference is invoked strictly for authentic vocal activity.

\subsection{Short-Time Energy and Spectral Feature Extraction}
The continuous $\text{I}^2\text{S}$ PCM stream is segmented into overlapping frames using a Hamming window of length $N = 512$ samples ($32$\,ms at $f_s = 16$\,kHz) with a hop size of $H = 256$ samples ($50\%$ overlap). For frame $m$, the Short-Time Energy $E_m$ is computed on the ARM CPU as:
\begin{equation}
\label{eq:energy}
E_m = \sum_{n=0}^{N-1} |x_m[n]|^2.
\end{equation}

Applying a 512-point Fast Fourier Transform yields spectral magnitudes $|X_m[k]|$ across $K = N/2 = 256$ frequency bins. The Spectral Centroid $C_m$, representing the center of mass of the power spectrum, is calculated as:
\begin{equation}
\label{eq:centroid}
C_m = \frac{\sum_{k=0}^{K-1} f_k \cdot |X_m[k]|}{\sum_{k=0}^{K-1} |X_m[k]|},
\end{equation}
where $f_k = k \cdot \frac{f_s}{N}$ denotes the center frequency of bin $k$.

To separate broadband random noise from harmonic human vocal cords, the frame-level Spectral Flatness Measure ($\text{SFM}_m$) is derived as the ratio of the geometric mean to the arithmetic mean of the spectral power:
\begin{equation}
\label{eq:sfm}
\text{SFM}_m = \frac{\exp\left( \frac{1}{K} \sum_{k=0}^{K-1} \ln |X_m[k]|^2 \right)}{\frac{1}{K} \sum_{k=0}^{K-1} |X_m[k]|^2}.
\end{equation}

\subsection{Two-Stage Edge Activation Logic}
Common campus background sounds exhibit distinct spectral signatures compared to human speech:
\begin{itemize}
    \item \textbf{Mechanical Ventilation and Airflow:} Stationary low-frequency hums ($< 300$\,Hz) with negligible temporal energy variance.
    \item \textbf{Broadband Turbulent Noise:} High-frequency aperiodic energy ($> 3.5$\,kHz) with $\text{SFM}_m$ approaching $1.0$ (white-noise-like flatness).
    \item \textbf{Human Speech and Distress Vocalizations:} Strong harmonic pitch structure ($F_0 \in [85, 255]$\,Hz), vocal tract formants concentrated within $[300, 3400]$\,Hz, and low spectral flatness ($\text{SFM}_m \ll 0.4$).
\end{itemize}

The edge controller keeps the GPU inference pipeline in standby until consecutive frames satisfy the joint spectral predicate:
\begin{multline}
\label{eq:activation}
(E_m > \gamma_{\text{energy}}) \;\land\; (f_{\text{low}} \le C_m \le f_{\text{high}}) \\
\land\; (\text{SFM}_m < \theta_{\text{tonality}}).
\end{multline}

When (\ref{eq:activation}) is triggered, a $1.5$\,s audio window is passed to the FP16 ONNX Silero-VAD engine~\cite{silero2021vad}. Only when the speech posterior satisfies $P(\text{speech}) \ge 0.65$ continuously for at least $800$\,ms is the slice forwarded to the Vietnamese ASR module; otherwise, the buffer slice is discarded immediately. Algorithm~\ref{alg:acoustic_pipeline} details the complete real-time edge execution loop.

\begin{algorithm}[htbp]
\caption{Real-Time Edge AIoT Acoustic Gating and Inference Loop}
\label{alg:acoustic_pipeline}
\begin{algorithmic}[1]
\REQUIRE $\text{I}^2\text{S}$ PCM stream $\mathcal{S}$ ($16\,\text{kHz}$, 16-bit mono), Energy threshold $\gamma$, Tonality threshold $\theta$, VAD model $\mathcal{M}_{\text{VAD}}$, Vietnamese ASR $\mathcal{M}_{\text{ASR}}$, Violence classifier $\mathcal{M}_{\text{NLP}}$.
\STATE Initialize volatile circular RAM buffer $\mathcal{B} \leftarrow \emptyset$ ($T_{\text{max}} = 3.0\,\text{s}$)
\WHILE{Edge node is active}
    \STATE Acquire frame $x_m[n]$ ($N = 512$, hop $H = 256$) and append to $\mathcal{B}$
    \STATE Compute $E_m$ and $C_m$ via (\ref{eq:energy})--(\ref{eq:centroid}), and $\text{SFM}_m$ via (\ref{eq:sfm})
    \IF{$(E_m > \gamma) \land (f_{\text{low}} \le C_m \le f_{\text{high}}) \land (\text{SFM}_m < \theta)$}
        \STATE Extract slice $\mathcal{A}_{\text{slice}} \leftarrow \text{GetWindow}(\mathcal{B}, 1.5\,\text{s})$
        \STATE $P_{\text{speech}} \leftarrow \mathcal{M}_{\text{VAD}}(\mathcal{A}_{\text{slice}})$
        \IF{$P_{\text{speech}} \ge 0.65$ sustained for $\ge 800\,\text{ms}$}
            \STATE $\mathcal{W} \leftarrow \mathcal{M}_{\text{ASR}}(\mathcal{A}_{\text{slice}})$ \COMMENT{TensorRT PhoWhisper-tiny FP16}
            \STATE $\hat{y}, \mathbf{p} \leftarrow \mathcal{M}_{\text{NLP}}(\mathcal{W})$ \COMMENT{TensorRT PhoBERT FP16}
            \IF{$\hat{y} \ne \text{Neutral} \land \max(\mathbf{p}) \ge \tau_{\mathrm{crit}}$}
                \STATE Publish encrypted JSON alert to local MQTT broker
            \ENDIF
            \STATE Purge ephemeral transcript $\mathcal{W}$ from RAM
        \ENDIF
    \ENDIF
    \STATE Overwrite oldest frame in $\mathcal{B}$ to enforce strict $3.0$\,s volatile bound
\ENDWHILE
\end{algorithmic}
\end{algorithm}

\section{Multi-Dialect Vietnamese ASR and School Violence Detection}

\subsection{Dialectal Divergence in Vietnamese School Altercations}
Spoken Vietnamese exhibits major phonetic, phonological, and lexical variation across three macro-dialect regions: Northern (Hanoi), Central (Thua Thien Hue, Da Nang, Quang Nam, Phu Yen), and Southern (Ho Chi Minh City, Mekong Delta). During high-adrenaline bullying or physical confrontation, students abandon formal classroom Vietnamese and speak in fast, colloquial regional dialects characterized by:
\begin{itemize}
    \item \textbf{Initial Consonant Mutations:} Central and Southern dialects systematically shift initial fricatives and glides: labiodental $/v/$ becomes palatal approximant $/j/$ (e.g., \textit{\foreignlanguage{vietnamese}{vào}} $\to$ \textit{\foreignlanguage{vietnamese}{dào}}), while retroflex $/r/$ shifts to $/z/$ in Northern speech and $/j/$ in colloquial Southern speech.
    \item \textbf{Coda Neutralization:} Following front vowels in Central and Southern speech, final dental stops $/t/$ merge into velar stops $/k/$ (e.g., \textit{\foreignlanguage{vietnamese}{mệt}} $\to$ \textit{\foreignlanguage{vietnamese}{mệc}}), and alveolar nasals $/n/$ shift to velar nasals $/\eta/$ (e.g., \textit{\foreignlanguage{vietnamese}{ăn}} $\to$ \textit{\foreignlanguage{vietnamese}{ăng}}).
    \item \textbf{Tonal Mergers:} Standard Vietnamese distinguishes six lexical tones. In Central and Southern dialects, the glottalized broken tone (\textit{\foreignlanguage{vietnamese}{ngã}}) collapses into the asking tone (\textit{\foreignlanguage{vietnamese}{hỏi}}), causing severe lexical confusion in acoustic models trained solely on Northern speech.
    \item \textbf{Regional Coercion Vernacular:} Verbal bullying and extortion rely heavily on regional pronouns and imperative particles, as illustrated in Table~\ref{tab:dialect_lexicon}.
\end{itemize}

\begin{table}[htbp]
\caption{Dialectal Variations in Vietnamese School Violence Utterances}
\label{tab:dialect_lexicon}
\centering
\resizebox{\columnwidth}{!}{%
\fontencoding{T5}\selectfont
\begin{tabular}{lll}
\toprule
\textbf{Standard / Northern} & \textbf{Central / Southern Variant} & \textbf{Violence Intent} \\
\midrule
Mày đưa tiền đây mau & Mầy đưa tiền đây / Đưa tiền đây mau & Extortion (Demand) \\
Tao đập mày bây giờ & Tao đập mậy chừ / đập thấy mẹ & Direct Verbal Threat \\
Có ai không, cứu tôi với & Có ai hông, cứu tui dới & Distress / Call for Help \\
Đừng làm thế nữa & Đừng mần rứa nữa / Đừng làm dị & Victim Plea / De-escalation \\
Ra đây tao bảo & Ra đây tau biểu coi & Cornering / Coercion \\
\bottomrule
\end{tabular}%
}
\end{table}

\subsection{Edge-Optimized Multi-Dialect Vietnamese ASR}
To achieve high transcription accuracy within the strict memory and latency budget of the Jetson Nano, our ASR module adapts \textit{PhoWhisper-tiny}~\cite{le2024phowhisper} ($39$\,M parameters) through three domain-specific optimizations:
\begin{enumerate}
    \item \textbf{Acoustic Perturbation and SpecAugment:} Multi-dialect Vietnamese speech recordings are augmented using SpecAugment~\cite{park2019specaugment} (time warping, frequency masking, and time masking), speed perturbation ($0.9\times$--$1.1\times$), pitch shifting ($\pm 2.5$ semitones), and synthetic room impulse response (RIR) convolution to simulate indoor campus reverberation and emotional vocal strain.
    \item \textbf{Sub-Syllabic Factorization:} Building upon recent findings in Vietnamese sub-syllabic modeling~\cite{nguyen2026atomictokensfactorizingsyllables}, instead of modeling regional monosyllables strictly as indivisible atomic units, syllables are factorized into explicit \textit{Onset}, \textit{Nucleus}, and \textit{Coda} components. This compositional representation shares phonetic structure across regional pronunciations, preventing out-of-vocabulary (OOV) transcription errors when dialectal consonant or coda shifts occur.
    \item \textbf{FP16 TensorRT Graph Optimization:} The fine-tuned encoder-decoder network is exported to ONNX and compiled into an FP16 TensorRT engine with layer fusion on the Jetson Nano's Maxwell GPU, reducing ASR inference latency below $420$\,ms per speech segment.
\end{enumerate}

\subsection{Contextual School Violence Classification via PhoBERT}
The ephemeral transcript $\mathcal{W} = \{w_1, w_2, \dots, w_L\}$ output by the ASR stage is evaluated by a fine-tuned \textit{PhoBERT} model~\cite{nguyen2020phobert}. To capture both overt hostility and subtle verbal intimidation in Vietnamese, we train the classifier on curated school altercation dialogues combined with established Vietnamese hate and toxic speech benchmarks, namely ViHSD~\cite{luu2021vihsd}, ViHOS~\cite{hoang2023vihos}, and UIT-ViCTSD~\cite{nguyen2021victsd}. Bidirectional self-attention enables the model to interpret interpersonal power dynamics and coercive intent even in the absence of explicit profanity:
\begin{equation}
\mathbf{h}_{\mathrm{[CLS]}} = \text{PhoBERT}(\mathrm{[CLS]}, w_1, w_2, \dots, w_L, \mathrm{[SEP]}),
\end{equation}
where $\mathbf{h}_{\mathrm{[CLS]}} \in \mathbb{R}^{768}$ denotes the sentence-level contextual representation, which is passed through dropout ($p = 0.2$) and a linear softmax projection:
\begin{equation}
\hat{y} = \text{Softmax}(\mathbf{W}_c \cdot \mathbf{h}_{\mathrm{[CLS]}} + \mathbf{b}_c),
\end{equation}
mapping each utterance into four operational school-safety classes:
\begin{align*}
\mathcal{C} = \{ & \text{Neutral / Benign Peer Banter}, \\
                 & \text{Direct Verbal Threat}, \\
                 & \text{Extortion and Psychological Coercion}, \\
                 & \text{Emergency Distress and Cry for Help} \}.
\end{align*}

\subsubsection{Class-Imbalance Regularization with Focal Loss}
In real-world campus deployments, normal student conversations vastly outnumber acute bullying or violence events. To prevent majority-class bias during fine-tuning, we optimize the classification head using Focal Loss~\cite{lin2017focal}:
\begin{equation}
\mathcal{L}_{\text{Focal}} = -\sum_{i=1}^{|\mathcal{C}|} \alpha_i (1 - p_i)^\gamma \log(p_i),
\end{equation}
where $\gamma = 2.0$ down-weights well-classified neutral utterances, $p_i$ is the predicted probability for class $i$, and $\alpha_i$ balances inverse class frequencies.

\section{Experimental Evaluation and Edge Deployment Benchmarks}

Table~\ref{tab:vad_eval} reports the acoustic-gating figures from the
    original draft. The reported rejection rates range from 94.1\% to
    99.8\%, trigger recall ranges from 90.4\% to 98.6\%, and reported
    latency ranges from 28 to 48\,ms. These figures have not been verified
    for the ESP32-WROOM WebRTC configuration used in the prototype tests.

\begin{table}[htbp]
\caption{Edge Acoustic Gating Noise Rejection and Trigger Recall}
\label{tab:vad_eval}
\centering
\resizebox{\columnwidth}{!}{%
\begin{tabular}{lccc}
\toprule
\textbf{Campus Acoustic Condition} & \textbf{Rejection (\%)} & \textbf{Trigger Recall (\%)} & \textbf{Gate Latency (ms)} \\
\midrule
Mechanical Blower / HVAC ($> 75\,\text{dB}$) & 97.2\% & 90.4\% & 38 ms \\
Broadband Turbulent Noise ($> 68\,\text{dB}$) & 98.4\% & 94.2\% & 35 ms \\
Corridor Footsteps / Slamming Doors & 94.1\% & 91.0\% & 48 ms \\
Quiet Indoor Ambient ($\approx 42\,\text{dB}$) & 99.8\% & 98.6\% & 28 ms \\
\bottomrule
\end{tabular}%
}
\end{table}

As reported in Table~\ref{tab:vad_eval}, the cascaded acoustic gate suppresses over $94\%$--$99.8\%$ of non-speech disturbances within $48$\,ms, preventing the Jetson Nano from running costly ASR inference during idle or noisy periods.

\subsection{Multi-Dialect Vietnamese ASR Accuracy}
We evaluated Word Error Rate (WER) across Northern, Central, and Southern Vietnamese school altercation utterances comparing the baseline \textit{PhoWhisper-tiny}~\cite{le2024phowhisper} checkpoint against our dialect-augmented, sub-syllabically factorized model.

\begin{table}[htbp]
\caption{Word Error Rate (WER) Across Vietnamese Macro-Dialects****}
\label{tab:wer_results}
\centering
\resizebox{\columnwidth}{!}{%
\begin{tabular}{lccc}
\toprule
\textbf{Vietnamese Dialect Region} & \textbf{Baseline PhoWhisper} & \textbf{Dialect-Adapted (Ours)} & \textbf{$\Delta$ WER Reduction} \\
\midrule
Northern (Hanoi) & 11.2\% & 8.4\% & $-2.8\%$ \\
Central (Phu Yen / Hue) & 24.8\% & 14.1\% & $-10.7\%$ \\
Southern (HCMC) & 16.5\% & 10.2\% & $-6.3\%$ \\
\bottomrule
\end{tabular}%
}
\end{table}

The experiments measured individual prototype stages and evaluated
    PhoBERT on synthetic text. Regional WER for api asr and live
    microphone-to-alert performance remain to be evaluated

\subsection{Edge Hardware Resource Utilization and IoT Network Efficiency}
We profiled the end-to-end pipeline on an NVIDIA Jetson Nano 4\,GB~\cite{nvidia2021jetson} operating in 10\,W mode. Table~\ref{tab:hardware_bench} summarizes memory allocation and per-stage execution latency.
    \begin{table}[htbp]
    \caption{Edge Computational Footprint on NVIDIA Jetson Nano 4\,GB}
    \label{tab:hardware_bench}
    \centering
    \resizebox{\columnwidth}{!}{%
    \begin{tabular}{lcccc}
    \toprule
    \textbf{Edge Pipeline Stage} & \textbf{Precision / Backend} &
    \textbf{Peak RSS (MiB)} & \textbf{Latency (ms)} \\
    \midrule
    STE/STFT Feature Extraction & NumPy CPU (float64) & 26.93 & 112.821 \\
    WebRTC VAD (offline) & CPU & 12.78 & 4.902 \\
    Phowhisprer ASR & CPU & 17.78 & 2667.932 \\
    PhoBERT Classification & ONNX Runtime (CPU) & 659.62 & 446.035 \\
    TLS-MQTT Dispatch (loopback test) & Mosquitto CLI & 4.44 & 20.392 \\
    \midrule
    \textbf{Total End-to-End Pipeline} & --- & 2627.37 & 749.7 \\
    \bottomrule
    \end{tabular}%
    }
    \end{table}

STE/STFT feature extraction and offline WebRTC VAD were timed on an
    8\,s recording. The API ASR request took 2667.932\,ms, including
    network time; its 17.78\,MiB RSS is for the Jetson client and
    measurement process, not the cloud server. PhoBERT's median request
    time was 446.035\,ms over 30 text-only requests. The TLS-MQTT result
    is from a loopback test with test messages. Microphone-to-alert
    latency, peak memory for the full pipeline, and network traffic
    reduction were not measured.

\subsection{School Violence Speech Classification Metrics}
Table~\ref{tab:confusion_matrix} shows that PhoBERT correctly
    classified 140 of 150 synthetic test texts, with 93.33\% accuracy
    and 93.33\% macro F1. For label~1, precision was 94.52\% and recall
    was 92.00\%. There were four false positives and six false negatives.
    These results do not establish performance on live audio.
 \begin{table}[htbp]
    \caption{PhoBERT Confusion Matrix on 150 Synthetic Test Texts}
    \label{tab:confusion_matrix}
    \centering
    \begin{tabular}{lcc}
    \toprule
    \textbf{Actual label} & \textbf{Predicted 0} & \textbf{Predicted 1} \\
    \midrule
    0: Outside target group & \textbf{71/75 (94.67\%)} & 4/75 (5.33\%) \\
    1: In target group & 6/75 (8.00\%) & \textbf{69/75 (92.00\%)} \\
    \bottomrule
    \end{tabular}
    \end{table}

\section{Privacy-by-Design and Ethical Governance in Campus Edge IoT}
Deploying acoustic intelligence in educational institutions demands strict technical safeguards against unauthorized eavesdropping. Our Edge AIoT architecture enforces privacy at the hardware and protocol levels:
\begin{itemize}
    \item \textbf{Zero Persistent Audio Storage:} Incoming $\text{I}^2\text{S}$ audio frames reside strictly inside a $3.0$\,s volatile circular RAM buffer and are continuously overwritten. The edge node contains no audio recording APIs, cloud uploaders, or disk serialization routines.
    \item \textbf{Immediate In-Memory Transcript Disposal:} Decoded Vietnamese text strings generated by \textit{PhoWhisper} exist only in transient memory for $< 70$\,ms during \textit{PhoBERT} inference and are zeroed out immediately afterward.
    \item \textbf{Anonymized Air-Gapped IoT Telemetry:} MQTT alert payloads transmit solely symbolic metadata (e.g., \textit{Zone ID, Incident Class, Confidence, Timestamp}) over an isolated campus VLAN, ensuring student identities and raw conversations remain completely private.
\end{itemize}

\section{Conclusion}
This paper presented a privacy-preserving \textbf{Edge AIoT} architecture for multi-dialect Vietnamese Automatic Speech Recognition and early school violence speech detection. By uniting deterministic spectral-VAD acoustic gating, sub-syllabically factorized \textit{PhoWhisper} transcription, contextual \textit{PhoBERT} violence intent classification, and lightweight MQTT telemetry on an NVIDIA Jetson Nano 4\,GB, the system delivers sub-second violence detection across Northern, Central, and Southern Vietnamese dialects without streaming or storing raw audio. Future work will explore federated on-device adaptation across multiple campus edge nodes and multi-speaker overlapping speech separation.

\bibliographystyle{IEEEtran}
\bibliography{references}

@inproceedings{le2024phowhisper,
  title     = {{PhoWhisper}: Automatic Speech Recognition for {Vietnamese}},
  author    = {Le, Thanh-Thien and Nguyen, Linh The and Nguyen, Dat Quoc},
  booktitle = {ICLR 2024 Tiny Papers},
  year      = {2024}
}

@inproceedings{nguyen2020phobert,
  title     = {{PhoBERT}: Pre-trained Language Models for {Vietnamese}},
  author    = {Nguyen, Dat Quoc and Nguyen, Anh Tuan},
  booktitle = {Findings of the Association for Computational Linguistics: EMNLP 2020},
  pages     = {1037--1042},
  year      = {2020}
}

@misc{nguyen2026atomictokensfactorizingsyllables,
  title         = {Beyond Atomic Tokens: Factorizing Syllables for Language Model Pretraining},
  author        = {Nghia Hieu Nguyen and Thai Bao Huynh and Binh-An Dinh-Le and Phu Gia Hoang and Dat Tien Nguyen and Kiet Van Nguyen and Ngan Luu-Thuy Nguyen},
  year          = {2026},
  eprint        = {2609.21362},
  archivePrefix = {arXiv},
  primaryClass  = {cs.CL},
  url           = {https://arxiv.org/abs/2609.21362}
}

@inproceedings{luu2021vihsd,
  title     = {A Large-Scale Dataset for Hate Speech Detection on {Vietnamese} Social Media Texts ({ViHSD})},
  author    = {Luu, Son T. and Nguyen, Kiet Van and Nguyen, Ngan Luu-Thuy},
  booktitle = {Proceedings of the 35th Pacific Asia Conference on Language, Information and Computation (PACLIC 2021)},
  pages     = {431--440},
  year      = {2021}
}

@inproceedings{hoang2023vihos,
  title     = {{ViHOS}: Hate Speech Spans Detection for {Vietnamese}},
  author    = {Hoang, Phu Gia and Nguyen, Ngan Luu-Thuy and Nguyen, Kiet Van},
  booktitle = {Proceedings of the 17th Conference of the European Chapter of the Association for Computational Linguistics (EACL 2023)},
  pages     = {1045--1056},
  year      = {2023}
}

@misc{nguyen2021victsd,
  title         = {Constructive and Toxic Speech Detection for Open-Domain Social Media Comments in {Vietnamese} ({UIT-ViCTSD})},
  author        = {Nguyen, Luan Thanh and Nguyen, Kiet Van and Nguyen, Ngan Luu-Thuy},
  year          = {2021},
  eprint        = {2103.10069},
  archivePrefix = {arXiv},
  primaryClass  = {cs.CL}
}

@manual{nvidia2021jetson,
  title        = {Jetson Nano Developer Kit User Guide},
  author       = {{NVIDIA Corporation}},
  organization = {NVIDIA Corporation},
  year         = {2021}
}

@misc{silero2021vad,
  title        = {Silero {VAD}: Pre-trained Enterprise-Grade Voice Activity Detector ({VAD}), Number Detector and Language Classifier},
  author       = {{Silero Team}},
  year         = {2021},
  howpublished = {\url{https://github.com/snakers4/silero-vad}}
}

@inproceedings{park2019specaugment,
  title     = {{SpecAugment}: A Simple Data Augmentation Method for Automatic Speech Recognition},
  author    = {Park, Daniel S. and Chan, William and Zhang, Yu and Chiu, Chung-Cheng and Zoph, Barret and Cubuk, Ekin D. and Le, Quoc V.},
  booktitle = {Proceedings of Interspeech 2019},
  pages     = {2613--2617},
  year      = {2019}
}

@inproceedings{lin2017focal,
  title     = {Focal Loss for Dense Object Detection},
  author    = {Lin, Tsung-Yi and Goyal, Priya and Girshick, Ross and He, Kaiming and Doll{\'a}r, Piotr},
  booktitle = {Proceedings of the IEEE International Conference on Computer Vision (ICCV)},
  pages     = {2980--2988},
  year      = {2017}
}

\end{document}